\documentstyle[sprocl]{article}

\def\be{\begin{equation}}
\def\ee{\end{equation}}
\def\bea{\begin{eqnarray}}
\def\eea{\end{eqnarray}}
\begin{document}

\title{TRANSVERSE SPIN / TRANSVERSE MOMENTUM ANALOGY AND GAUGE INVARIANCE
OF DEEPLY VIRTUAL COMPTON SCATTERING}

\author{B. PIRE}

\address{
CPhT\footnote {Unit\'e Mixte de Recherche C7644 du Centre
National de la Recherche Scientifique}, \'Ecole Polytechnique,\\ F-91128
Palaiseau, France\\E-mail: pire@cpht.polytechnique.fr}

\author{\underline{O. V. TERYAEV}}

\address{Bogoliubov Laboratory of Theoretical Physics,JINR,\\
Dubna 141980, Russia\\E-mail: teryaev@thsun1.jinr.ru}

\maketitle\abstracts{ Using the similarity between twist 3 contributions
to transverse spin for usual parton distributions and transverse
momentum  for non-diagonal parton distributions, we determine
for the  simple case of a scalar (pion, $He^4$) target
the linear in transverse momentum twist-3 contribution
to Deeply Virtual Compton amplitude. This allows to write a gauge
invariant amplitude for this non-forward process.}

\vspace{.5cm}
Deeply Virtual Compton Scattering (DVCS) is a new hard process which
recently attracted  much attention\cite{Ji,OFPD}. It gives access to
new hadronic quantities which may turn to be important for describing the
physics of confinement: the  non-diagonal (off-forward, skewed)
parton distributions. It is known that the DVCS amplitude
is not manifestly gauge invariant, the violation coming as an effect
of non-zero transverse momentum $\Delta_T$ between the ingoing and
outgoing protons. We show here that a fruitful analogy between transverse
spin effects in usual deep inelastic scattering on the one hand and
transverse momentum effects in DVCS on the other hand helps to resolve
this problem.

Let us first unravel the physical nature of the problem. Gauge
invariance is intimately related to the possibility of expressing the
Lagrangian in terms of covariant derivatives. These derivatives mix usual
momentum and the gauge field. It is thus in general impossible to
separate transverse momentum effects from an additionnal gluon
interaction. In terms of Operator Product Expansion, parton transverse
momentum effects are at the same level as operators
containing an additionnal gluon to the minimum number of quark
fields. Thus only the sum of quark and quark gluon contributions is
gauge invariant.
The situation is well known in the case of
transverse polarization for deep inelastic scattering, {\em  i.e.
forward} virtual Compton scattering \cite{ET84}. The
 gauge invariant result appears only when both quark
and quark gluon matrix elements are taken into account. They are
combining the two sources of transverse spin effects, namely,
intrinsic transverse momentum and the  transverse gluon field, to a
gauge invariant structure, related to the covariant derivative.
By making use of the equations of motion the transverse spin-dependent
part of DIS hadronic tensor may be written in the manifestly gauge
invariant form, depending on the single matrix element $c_T(x)$:
\begin{eqnarray}
\label{def}
W^{\mu \nu} = \frac{iM}{pq} g_T (x_B) \epsilon^{\mu \nu \alpha
\beta} q^\alpha s_T^\beta, \\ \nonumber
g_T(x)=g_1(x)+g_2(x) = {1\over 2} (c_T(x)+c_T(-x)),   \\ \nonumber
2 M s_\mu c_T (x)=\int{d\lambda }e^{i\lambda x}
\langle p,s|\bar \psi (0) \gamma_\mu \gamma ^5\psi (\lambda n)|p,s \rangle
\end{eqnarray}
Let us now turn our attention to the off-diagonal case.
The dependence of the matrix element
$<p+\Delta/2|O|p-\Delta/2>$ is  similar to the dependence on the
transverse polarization.   The similarity between the forward matrix
element of the polarized  particle, and the
non-forward matrix element comes form the fact, that
in both cases one has a dependence on an extra space-like vector,
orthogonal to the particle momentum (provided the
latter is chosen in a symmetric way \cite{Ji}).
Such a procedure is valid for scalar or
longitudinally polarized particles.
For spin $1/2$ particles, the matrix elements
depend in principle of two vectors, the spin and the transverse
momentum
and the analogy with the
forward case should be worked out with the tensor polarization of spin
$1$ particles.
In the case of longitudinal spin, the spin vector is not
independent from the momentum and we may treat the matrix element as in
the scalar particle case.

There are however two important distinctions. First, the dependence
on the transverse polarization is linear, like for any density matrix,
while this is definitely not the case for the dependence on $\Delta_T$.
Second, the polarization is a pseudovector, contrarily to the transverse
momentum.
To meet with the first point we restrict ourselves to the
first non-vanishing term, linear in $\Delta_T$. The second point is a
mere technicality, as now the matrix elements containing Dirac matrices
$\gamma_\mu$ and $\gamma_\mu \gamma_5$ just interchange their kinematical
structure. The extra $\gamma_5$ factor may be attributed to the one of
the photon vertices, so that the {\it electromagnetic} contribution
to the DVCS is related to parity violating term\cite{Blum}
in the forward case.
The twist-3 contribution to the latter is known to be expressed
\cite{Ji2} in terms of  the same matrix element ($c_T$) as
appearing in the parity conserving case (\ref{def})
\begin{eqnarray}
\label{pv}
W^{\mu \nu} = \frac{M}{pq} (c_T(x_B)-c_T(-x_B))
[\tilde p^\mu s_T^\nu+
\tilde p^\nu s_T^\mu], \ \
\tilde p^\mu = p^\mu-q^\mu \frac{pq}{q^2}.
\end{eqnarray}
The expression for the off-diagonal hadronic tensor keeping only
linear terms in $\Delta_T$ is then
explicitely gauge invariant
\begin{eqnarray}
\label{fin}
H^{\mu \nu} = \frac{M}{pq} \int_0^1 dx c_\Delta(x) (\frac{1}{x-\xi}
+\frac{1}{x+\xi})
[\tilde p^\mu \Delta_T^\nu+
\tilde p^\nu \Delta_T^\mu], \\ \nonumber
2 M \Delta_T^\mu c_\Delta (x)=
\int{d\lambda }e^{i\lambda x}
\langle p+\Delta/2|\bar \psi (0) \gamma^\mu \psi (\lambda n)
|p-\Delta/2 \rangle.
\end{eqnarray}
Another proposal has been done\cite{gui} to correct the
DVCS amplitude by adding
a term proportionnal to
$\Delta_T$.
%\begin{equation}
%H_{DVCS}^{\mu\nu} \rightarrow H_{DVCS}^{\mu\nu} +
%p^{\mu}(\Delta_T)_\lambda H_{DVCS}^{\lambda\nu}
%\end{equation}
This is different from our proposed definition
as their result is governed by the same distribution $H$,
whether $\Delta_T$ vanishes or not. At the same time, our result
(\ref{fin}) reveals the kinematical
structure of $\Delta_T=0$ case (the appearing tensor structure
is actually the linear in $\Delta_T$ part of the tensor
$\tilde p^\mu  \tilde p^\nu$, whose coefficient is just  the
standard $F_2$ structure function), but multiplied by the
different long-distance matrix element. Here the analogy with the
transverse polarization is also clear.
The kinematical structure of the hadronic tensor for longitudinal
and transverse polarization is the same $
\epsilon^{\mu \nu \alpha
\beta} q^\alpha s^\beta$, while their coefficients
differ by the structure function $g_2$. Such a difference is
manifested because of the light cone direction violating the
rotational invariance, which is restored for the local operators
for the first moments, resulting in the Burkhardt-Cottingham sum rule.
We expect the similar sum rule to be valid in the non-diagonal case.

We are thankful to M. Diehl for valuable comments.
O.T. is indebted to A.V. Radyushkin for enlightening discussion
on the difference between longitudinal and transverse momentum
transfer in DVCS.

\end{document}